\documentclass[aps,pra,twocolumn,amsfonts,amssymb,amsmath,showpacs,
floatfix,nofootinbib,groupedaddress,superscriptaddress,citesort]{revtex4}
\usepackage{mathrsfs}
\usepackage{amsfonts}
\usepackage{amstext}
\usepackage{amsmath}
\usepackage{amssymb}
\usepackage{bm}
\usepackage{bbm}
\usepackage[dvips]{graphicx}
\def\qed{\leavevmode\unskip\penalty9999 \hbox{}\nobreak\hfill
     \quad\hbox{\leavevmode  \hbox to.77778em{%
              \hfil\vrule   \vbox to.675em%
               {\hrule width.6em\vfil\hrule}\vrule\hfil}}
     \par\vskip3pt}

\begin{document}

\title{Unextendible entangled bases with fixed Schmidt number\\}

\author{Yu Guo}
\email{guoyu3@aliyun.com}
\affiliation{School of Mathematics and Computer Science, Shanxi Datong University, Datong, Shanxi 037009, China}%

\author{Shengjun Wu}
\email{sjwu@nju.edu.cn}
\affiliation{Kuang Yaming Honors School, Nanjing University, Nanjing, Jiangsu 210093, China}

\begin{abstract}

The unextendible product basis (UPB) is generalized to the unextendible entangled basis with any arbitrarily given Schmidt number $k$ (UEBk)
for any bipartite system $\mathbb{C}^d\otimes\mathbb{C}^{d'}$ ($2\leq k<d\leq d'$),
which can also be regarded as a generalization of the unextendible maximally entangled basis (UMEB).
A general way of constructing such a basis with arbitrary $d$ and $d'$ is proposed.
Consequently, it is shown that there are at least $k-r$ (here $r=d$ mod $k$, or $r=d'$ mod $k$) sets of UEBk when $d$ or $d'$ is not the multiple of $k$, while
there are at least $2(k-1)$ sets of UEBk when both $d$ and $d'$ are the multiples of $k$.

\end{abstract}

\maketitle

\section{Introduction}

Entanglement between particles is a fundamental feature
of quantum physics and has been extensively investigated
in recent years \cite{Horodecki2009,Guhne}.
It is the key to understanding the deepest implications of
quantum mechanics to information theory and even to the
nature of reality \cite{Bell}. It was found that there are
sets of product states which nevertheless display a form of
nonlocality \cite{Bennett1999,Divincenzo}.

In Ref. \cite{Bennett1999}, the unextendible product
basis (UPB) is proposed, it is shown that
there are sets of
incomplete orthogonal product bases whose complementary space
does not contain product states.
The members of a UPB are not perfectly distinguishable by local positive
operator valued measurements and classical communication. In addition, UPB can be used for
constructing bound entangled states \cite{Pittenger}.

The notion of UPB is generalized to an unextendible maximally entangled basis (UMEB) in Ref.~\cite{Bravyi} by Bravyi and Smolin.
A UMEB is a set of orthonormal maximally entangled states in a two-qudit system consisting of fewer than
$d^2$ vectors which have no additional maximally entangled vectors orthogonal to all of them.
It is shown that there is no UMEB in the two-qubit system, while a six-member UMEB in $\mathbb{C}^3\otimes \mathbb{C}^3$ and
a 12-member UMEB in $\mathbb{C}^4\otimes \mathbb{C}^4$ were constructed.
Later, B. Chen and S.-M. Fei proved in Ref. \cite{Chen} that there exists a set of $d^2$-member UMEB in $\mathbb{C}^d\otimes \mathbb{C}^{d'}$ ($\frac{d'}{2}<d<d'$) and
questioned the existence of UMEBs in the case of $d'\geq 2d$. Very recently,
this problem is solved in Ref. \cite{Limaosheng}: there might be two sets of UMEBs in any bipartite system, and
an explicit construction of UMEBs is proposed.

One of the most indispensable quantity associated with bipartite basis is the \emph{Schmidt number}.
It can be used to
characterize and quantify the degree of bipartite entanglement for
pure state directly~\cite{Donald,Sperling} and, also, it can be
operationally interpreted as the zero-error entanglement cost in the
protocol of one-shot entanglement dilution~\cite{Buscemi}.
Note that UPB is a set of pure states with the minimal Schmidt number 1 and the UMEB is a set of pure
states with the maximal Schmidt number $d$.
Then, what about the case for unextendible entangled basis with Schmidt number $k$ ($2\leq k<\min\{d,d'\}$)?
We will investigate this problem in this short paper.
Consequently, we find out that there are no entangled states with Schmidt number greater than or equal to $k$ in the
complementary space of the constructed UEBk by considering
their Schmidt number. This implies that we can construct
entangled states with limited Schmidt number \cite{Schmidtrankofmixedstate} via the special
structure of the UEBk.

The material in this paper is arranged as follows. In
Sec. II we introduce the concept of UEBk. Section III puts forward a method of
constructing UEBk in $\mathbb{C}^d\otimes\mathbb{C}^{d'}$ system when $d'$
is not a multiple of $k$, where $k$ is an
arbitrarily given Schmidt number. Section IV
deals with the case for $d'$ is a multiple of $k$.
Finally,
we conclude in Sec. V.

\section{Definition}

For the sake of clarity, we recall the definition of UMEB first.

{\it Definition \cite{Chen}.} A set of states $\{|\phi_i\rangle\in\mathbb{C}^d\otimes\mathbb{C}^{d'}: i=1,2,\dots,m,m<dd'\}$ is called an
m-number UMEB if and only if

(i) $|\phi_i\rangle$, $i=1,2,\dots,m$, are maximally entangled;

(ii) $\langle\phi_i|\phi_j\rangle=\delta_{ij}$;

(iii) if $\langle\phi_i|\psi\rangle=0$ for all $i=1,2,\dots,m$, then $|\psi\rangle$ cannot be maximally entangled.

Here state $|\psi\rangle\in\mathbb{C}^d\otimes\mathbb{C}^{d'}$ is called a maximally entangled state if
it can be written as $|\psi\rangle=\frac{1}{\sqrt{d}}\sum_{i=0}^{d-1}|i\rangle|i'\rangle$ for some orthonormal basis
$\{|i\rangle\}$ of $\mathbb{C}^d$ and some orthonormal set $\{|i'\rangle\}$ of $\mathbb{C}^{d'}$.

With the same sprit, we define the UEBk. We denote by $S_r(|\psi\rangle)$ the Schmidt number of the pure state $|\psi\rangle\in\mathbb{C}^d\otimes\mathbb{C}^{d'}$.
Recall that, the Schmidt number of a pure state $|\psi\rangle\in\mathbb{C}^d\otimes\mathbb{C}^{d'}$ is defined
as the length of the Schmidt decomposition \cite{Schmidt}:
if
$|\psi\rangle=\sum_{k=0}^{m-1}\lambda_{k}|k\rangle|k'\rangle$ is its
Schmidt decomposition, then
$S_r(|\psi\rangle)=m$. It is clear that
$S_r(|\psi\rangle)={\rm rank}(\rho_1)={\rm rank}(\rho_2)$,
where $\rho_i$ denotes the reduced state of the $i-$th part.

{\it Definition.} A set of states $\{|\phi_i\rangle\in\mathbb{C}^d\otimes\mathbb{C}^{d'}: i=1,2,\dots,m,m<dd'\}$ is called an
m-number unextendible entangled bases with Schmidt number $k$ (UEBk) if and only if

(i) $S_r(|\phi_i\rangle)=k$, $i=1,2,\dots,m$;

(ii) $\langle\phi_i|\phi_j\rangle=\delta_{ij}$;

(iii) if $\langle\phi_i|\psi\rangle=0$ for all $i=1,2,\dots,m$, then $S_r(|\psi\rangle)\neq k$.

It is clear that UEBk reduces to UPB (UMEB) when $k=1$ ($k=d$).
In the following, we will show that UEBk exists in any bipartite system $\mathbb{C}^{d}\otimes\mathbb{C}^{d'}$ with $d>2$.
Hereafter, we always assume that $2\leq k<d\leq d'$.

\section{The case for $d'$
is not a multiple of $k$}

We first consider the case for $d'=tk+r$, $0<r<k$, where $k$ is an arbitrarily given Schmidt number.

{\it Proposition 1.} Let
\begin{eqnarray}
|\phi_{mnl}\rangle:=\frac{1}{\sqrt{k}}\sum\limits_{p=0}^{k-1}\zeta_k^{np}|p\oplus m\rangle|[(l-1)k+p]'\rangle,
\end{eqnarray}
where $m=0$, $1$, $\dots$, $d-1$, $n=0$, 1, $\dots$, $k-1$, $1<k<d$, $l=1$, 2, $\dots$, $t$,
$\zeta_k=e^\frac{2\pi\sqrt{-1}}{k}$ and $d'=tk+r$, $0< r<k$, $x\oplus m$ denotes $x+m$ mod $d$. Then $\{|\phi_{mnl}\rangle\}$ is a $tkd$-member UEBk in $\mathbb{C}^d\otimes\mathbb{C}^{d'}$.

{\it Proof.} (i) It is clear that $S_r(|\phi_{mnl}\rangle)=k$ for any $m$, $n$ and $l$.

(ii) Orthogonality.
\begin{eqnarray*}
&&\langle\phi_{\tilde{m}\tilde{n}\tilde{l}}|\phi_{mnl}\rangle\nonumber\\
&=&\frac{1}{k}\sum\limits_{p=0}^{k-1}\sum\limits_{\tilde{p}=0}^{k-1}
\zeta_k^{np-\tilde{n}\tilde{p}}\langle \tilde{p}\oplus \tilde{m}|p\oplus m\rangle\nonumber\\
&&\times\langle[(\tilde{l}-1)k+\tilde{p}]'|[(l-1)k+p]'\rangle\nonumber\\
&=&\frac{1}{k}\sum\limits_{p=0}^{k-1}\zeta_k^{(n-\tilde{n})p}\langle p\oplus \tilde{m}|p\oplus m\rangle\delta_{l\tilde{l}}\nonumber\\
&=&\delta_{m\tilde{m}}\delta_{n\tilde{n}}\delta_{l\tilde{l}}.
\end{eqnarray*}

(iii) Unextendibility.
Let $V_1$ denote the subspace spanned by
\begin{eqnarray*}
\{|\phi_{mnl}\rangle:m=0,\dots,d-1,~~ n=0,\dots,k-1,~~\\
1<k<d,~~l=1,\dots,t\}.
\end{eqnarray*}
Then $\dim V_1=tkd$. One can easily check that any vector $|\psi\rangle\in V_1^{\perp}$
has the form
\begin{eqnarray*}
|\psi\rangle=\sum\limits_{i=0}^{d-1}\sum\limits_{j=0}^{r-1}a_{ij}|i\rangle|(tk+j)'\rangle.
\end{eqnarray*}
It turns out that $S_r(|\psi\rangle)<k$.
\hfill$\blacksquare$

Furthermore, we can formulate the following fact.

{\it Proposition 2.} Let
\begin{eqnarray}
|\phi_{mnl}\rangle:=\frac{1}{\sqrt{k}}\sum\limits_{p=0}^{k-1}\zeta_k^{np}|p\oplus m\rangle|[(l-1)k+p]'\rangle,
\end{eqnarray}
where $m=0$, $1$, $\dots$, $d-q-1$, $n=0$, 1, $\dots$, $k-1$, $1<k<d$, $l=1$, 2, $\dots$, $t$,
$\zeta_k=e^\frac{2\pi\sqrt{-1}}{k}$ and $d'=tk+r$, $0< r<k$, $x\oplus m$ denotes $x+m$ mod $d-k+q$ with $1\leq q<k-r$.
Then $\{|\phi_{mnl}\rangle\}$ is a $(d-q)tk$-member UEBk in $\mathbb{C}^d\otimes\mathbb{C}^{d'}$.

{\it Proof.} We only need to check the unextendibility.
Let $V_1$ denote the subspace spanned by
\begin{eqnarray*}
\{|\phi_{mnl}\rangle:m=0,\dots,d-q-1,~~n=0,\dots,k-1,~~\\
1<k<d,~~l=1,\dots,t\}.
\end{eqnarray*}
Then $\dim V_1=tk(d-q)$. It is easy to see that any vector $|\psi\rangle\in V_1^{\perp}$
admits the form  of $|\psi\rangle=\alpha|\psi_1\rangle+\beta|\psi_2\rangle$, where
\begin{eqnarray*}
|\psi_1\rangle&=&\sum\limits_{i=0}^{d-q-1}\sum\limits_{j=0}^{r-1}a_{ij}^{(1)}|i\rangle|(tk+j)'\rangle,\\
|\psi_2\rangle&=&\sum\limits_{i=d-q}^{d-1}\sum\limits_{j=0}^{d'-1}a_{ij}^{(2)}|i\rangle|j'\rangle.
\end{eqnarray*}
It turns out that $S_r(|\psi\rangle)<k$.
\hfill$\blacksquare$

Proposition 1 and Proposition 2 imply that there are at leat $k-r$ sets of UEBk for any possible $k$.

The UEBk in Proposition 1 can not be extended from the one in Proposition 2 since the former takes mod $d$ while the
latter one takes mod $d-k+q$.

\section{The case for $d'$
is a multiple of $k$}

There are two different cases when $d'$
is a multiple of the Schmidt number: $d$ is not a multiple of the Schmidt number or $d$ is a multiple of the Schmidt number.
We discuss them respectively.

\subsection{ $d$
is not a multiple of $k$}

{\it Proposition 3.} If $d=sk+r$, $0<r<k$, and $d'=tk$, we let
\begin{eqnarray}
&|\phi_{ijmn}\rangle~~~~~~~~~~~~~~~~~~~~~~~~~~~~~~~~~~~~~~~~~~~~~~~~~\nonumber\\
&=\frac{1}{\sqrt{k}}\sum\limits_{p=0}^{k-1}\zeta_k^{np}|(i-1)k+p\rangle|(((j-1)k+p)\oplus m)'\rangle,
\end{eqnarray}
where $i=1,2,\dots,s$,
$j=1,2,\dots,t$,
$m,n=0,1,\dots, k-1$, $x\oplus m$ denotes $x+m$ mod $d'$.
Then $\{|\phi_{ijmn}\rangle\}$ is a $stk^2$-member UEBk in $\mathbb{C}^d\otimes\mathbb{C}^{d'}$.

{\it Proof.} (i) It is clear that $S_r(|\phi_{ijmn}\rangle)=k$ for any $i$, $j$, $m$ and $n$.

(ii) Orthogonality.
\begin{eqnarray*}
&&\langle\phi_{\tilde{i}\tilde{j}\tilde{m}\tilde{n}}|\phi_{ijmn}\rangle\\
&=&\frac{1}{k}\sum\limits_{p=0}^{k-1}\sum\limits_{\tilde{p}=0}^{k-1}
\zeta_k^{np-\tilde{n}\tilde{p}}\langle (\tilde{i}-1)k+\tilde{p}|(i-1)k+p\rangle\\
&&\times\langle (((\tilde{j}-1)k+\tilde{p})\oplus \tilde{m})'|(((j-1)k+p)\oplus m)'\rangle\\
&=&\frac{1}{k}\sum\limits_{p=0}^{k-1}\zeta_k^{(n-\tilde{n})p}\langle (\tilde{i}-1)k+p|(i-1)k+p\rangle\delta_{j\tilde{j}}\delta_{m\tilde{m}}\\
&=&\delta_{i\tilde{i}}\delta_{j\tilde{j}}\delta_{m\tilde{m}}\delta_{n\tilde{n}}.
\end{eqnarray*}

(iii) Unextendibility.
Let $V_1$ denote the subspace spanned by
\begin{eqnarray*}
\{|\phi_{ijmn}\rangle:i=1,\dots,s,
j=1,\dots,t,
m,n=0,\dots,k-1\}.
\end{eqnarray*}
Then $\dim V_1=stk^2$ and any vector $|\psi\rangle\in V_1^{\perp}$
can be written as
\begin{eqnarray*}
|\psi\rangle=\sum\limits_{i=0}^{r-1}\sum\limits_{j=0}^{d'}a_{ij}|i\rangle|j'\rangle.
\end{eqnarray*}
It follows that $S_r(|\psi\rangle)<k$.
\hfill$\blacksquare$

Similar to Proposition 2, we have the following result.

{\it Proposition 4.} If $d=sk+r$, $0<r<k$, and $d'=tk$, we let
\begin{eqnarray}
&|\phi_{ijmn}\rangle~~~~~~~~~~~~~~~~~~~~~~~~~~~~~~~~~~~~~~~~~~~~~~~~~\nonumber\\
&=\frac{1}{\sqrt{k}}\sum\limits_{p=0}^{k-1}\zeta_k^{np}|(i-1)k+p\rangle|(((j-1)k+p)\oplus m)'\rangle,
\end{eqnarray}
where $i=1,2,\dots,s$,
$j=1,2,\dots,t$,
$m,n=0,1,\dots, k-1$, $x\oplus m$ denotes $x+m$ mod $d'-k+q$ with $1\leq q<k-r$.
Then $\{|\phi_{ijmn}\rangle: m\neq q,q+1,\dots,k-1\ {\rm when}\ j=t\}$ is a $sk(tk-k+q)$-member UEBk in $\mathbb{C}^d\otimes\mathbb{C}^{d'}$.

That is there are at leat $k-r$ sets of UEBk for any possible $k$ in such a case.

\subsection{$d$
is a multiple of $k$}

We consider now the case for $d=sk$ and $d'=tk$.

{\it Proposition 5.} If $d=sk$ and $d'=tk$, we let
\begin{eqnarray}
&|\phi_{ijmn}\rangle~~~~~~~~~~~~~~~~~~~~~~~~~~~~~~~~~~~~~~~~~~~~~~~~~~\nonumber\\
&=\frac{1}{\sqrt{k}}\sum\limits_{p=0}^{k-1}\zeta_k^{np}|(i-1)k+p\rangle|(((j-1)k+p)\oplus m)'\rangle,
\end{eqnarray}
where $i=1,2,\dots,s$,
$j=1,2,\dots,t$,
$m,n=0,1,\dots, k-1$, $x\oplus m$ denotes $x+m$ mod $d'-k+q$ with $1\leq q<k$.
Then $\{|\phi_{ijmn}\rangle: m\neq q,q+1,\dots,k-1\ {\rm{when}}\ j=t\}$ is a $sk(tk-k+q)$-member UEBk in $\mathbb{C}^d\otimes\mathbb{C}^{d'}$.

{\it Proof.} We only need to prove the unextendibility.
Let $V_1$ denote the subspace spanned by
\begin{eqnarray*}
\{|\phi_{ijmn}\rangle:i=1,\dots,s,
j=1,\dots,t,
m,n=0,1,\dots, k-1\}.
\end{eqnarray*}
Then $\dim V_1=sk(tk-k+q)$. Any vector $|\psi\rangle\in V_1^{\perp}$
has the form
\begin{eqnarray*}
|\psi\rangle=\sum\limits_{i=0}^{d}\sum\limits_{j=0}^{q-1}a_{ij}|i\rangle|j'\rangle.
\end{eqnarray*}
That is $S_r(|\psi\rangle)<k$.
\hfill$\blacksquare$

Symmetrically, the following is true.

{\it Proposition 6.} If $d=sk$ and $d'=tk$, we let
\begin{eqnarray}
&|\phi_{ijmn}\rangle~~~~~~~~~~~~~~~~~~~~~~~~~~~~~~~~~~~~~~~~~~~~~~~~~\nonumber\\
&=\frac{1}{\sqrt{k}}\sum\limits_{p=0}^{k-1}\zeta_k^{np}|((i-1)k+p)\oplus m\rangle|((j-1)k+p)'\rangle,
\end{eqnarray}
where $i=1,2,\dots,s$,
$j=1,2,\dots,t$,
$m,n=0,1,\dots, k-1$, $x\oplus m$ denotes $x+m$ mod $d-k+q$ with $1\leq q<k$.
Then $\{|\phi_{ijmn}\rangle: m\neq q,q+1,\dots,k-1\ {\rm{when}}\ i=s\}$ is a $tk(sk-k+q)$-member UEBk in $\mathbb{C}^d\otimes\mathbb{C}^{d'}$.

We thus conclude that there are at least $2(k-1)$ sets of UEBk when both $d$ and $d^{\prime}$ are multiples of the Schmidt number $k$.


\section{Conclusion and discussion}

The notion of UEBk is put forward, which extends the concepts of both UPB and UMEB.
We show that UEBk exists in any bipartite systems with $d>2$, and we explicitly construct the UEBks for different cases.
So far the existence problem of unextendible basis is settled thoroughly:
the unextendible basis always exists in any bipartite system with $d>2$
(here, UPB, UMEB and UEBk are collectively called unextendible basis).

For the special case of $d=sk$, we can give another method of constructing UEBks. Inspired by proposition 2
in Ref.~\cite{Limaosheng}, if $d=sk$, for any integer
\begin{widetext}
\begin{eqnarray}
m\in\left\{\begin{array}{ll}
\{d'-1,d'-2,\dots,d'-k+1\}, & {\rm if}\ d'\geq 2d,\\
\{d'-1,d'-2,\dots,d'-r\}, &{\rm if}\ d<d'<2d,\ d'=tk+r,\ r<k,
\end{array}\right.
\end{eqnarray}
\end{widetext}
we let
\begin{eqnarray}
&|\phi_{ijn}\rangle~~~~~~~~~~~~~~~~~~~~~~~~~~~~~~~~~~~~~~~~~~~~~~~~~\nonumber\\
&=\frac{1}{\sqrt{k}}\sum\limits_{p=0}^{k-1}\zeta_k^{np}|(i-1)k+p\rangle|(((i-1)k+p)\oplus j)'\rangle,\label{d=sk}
\end{eqnarray}
where $i=1,2,\dots,s$,
$j=0,1,\dots,m-1$,
$n=0,1,\dots, k-1$, $x\oplus j$ denotes $x+j$ mod $m$.
Then $\{|\phi_{ijn}\rangle\}$ is a $smk$-member UEBk in $\mathbb{C}^d\otimes\mathbb{C}^{d'}$.
Proposition 2 in Ref. \cite{Limaosheng} can be obtained easily from UEBk in Eq.~(\ref{d=sk}) by setting $k=d$;
and Proposition 1 in Ref. \cite{Limaosheng} can also be obtained from our Proposition 1 by setting $k=d$.

In addition, for any $m$-member UEBk $\{|\phi_i\rangle\}$ obtained from our scenario, we know from the Proofs that $S_r(|\psi\rangle) < k$ if $\langle\phi_i|\psi\rangle=0$ for all $i=1,2,\dots,m$.
Therefore, the range of the state
\begin{eqnarray}
\rho^{\perp}=\frac{1}{dd'-m}(I-\sum\limits_{i=1}^{m}|\phi_i\rangle\langle\phi_i|),
\end{eqnarray}
has no state with Schmidt number greater than or equal to $k$, namely, the Schmidt number
of $\rho^{\perp}$ is smaller than $k$ \cite{Schmidtrankofmixedstate}.
That is, UEBk can be used for constructing entangled state with Schmidt number smaller than $k$.
In other words, based on our methods, any state with Schmidt number greater than or equal
to $k$ lives in the subspace spanned by the UEBk and vice versa.
Namely, the space $\mathbb{C}^d\otimes\mathbb{C}^{d'}$ is divided into two subspace for any possible Schmidt number $k$:
one contains states with Schmidt number greater than or equal to $k$ and the other contains only states with
Schmidt number smaller than $k$.
However, for any given mixed state, its Schmidt number is hard to calculate since it is defined via all its pure state ensembles \cite{Schmidtrankofmixedstate}.
Therefore, UEBk can be used to construct entangled states with limited Schmidt number, thus provides us a useful tool in studying entanglement and related problems.

\begin{acknowledgements}
The authors wish to dedicate this work to Prof. Jinchuan Hou on the occasion of his 60th birthday.
Y. Guo acknowledges support from the Natural Science Foundation of China (Grant No. 11301312, Grant No. 11171249),
the Natural Science Foundation of Shanxi
(Grant No. 2013021001-1,  Grant No. 2012011001-2)
and the Research start-up fund for Doctors of Shanxi Datong University
(Grant No. 2011-B-01).
S. Wu acknowledges support from the Natural Science Foundation of China (Grant No. 11275181)
and the Fundamental Research Funds for the Central Universities (Grant No. 20620140531).
\end{acknowledgements}



\begin{thebibliography} {99}


\bibitem{Horodecki2009} R. Horodecki, P. Horodecki, M. Horodecki, and K. Horodecki,
Rev. Mod. Phys. \textbf{81}, 865 (2009).


\bibitem{Guhne} O. G\"{u}hne and G. T\'{o}th,
Phys. Rep. \textbf{474}, 1 (2009).


\bibitem{Bell} J. S. Bell, Physics (NY) \textbf{1}, 195 (1964).

\bibitem{Bennett1999} C. H. Bennett, D. P. DiVincenzo, T. Mor, P. W. Shor,
J. A. Smolin, and B. M. Terhal, Phys. Rev. Lett. \textbf{82}, 5385
(1999).

\bibitem{Divincenzo} D. P. Divincenzo, T. Mor, P. W. Shor, J. A. Smolin,
and B. M. Terhal, Commun. Math. Phys. \textbf{238}, 379
(2003).


\bibitem{Pittenger} A. O. Pittenger,
Lin. Alg. Appl. \textbf{359},  235--248 (2003)










\bibitem{Bravyi} S. Bravyi and J. A. Smolin, Phys. Rev. A \textbf{84}, 042306
(2011).


\bibitem{Chen} B. Chen and S. M. Fei, Phys. Rev. A \textbf{88}, 034301 (2013).


\bibitem{Limaosheng} M.-S. Li, Y.-L. Wang, and Z.-J. Zheng, Phys. Rev. A \textbf{89}, 062313 (2014).


\bibitem{Donald} M. J. Donald, M. Horodecki, and O. Rudolph,
J. Math. Phys. \textbf{43}, 4252--4272 (2002).





\bibitem{Sperling} J. Sperling and W. Vogel,
Phys. Scr.
\textbf{83}, 045002 (2011).



\bibitem{Buscemi} F. Buscemi and N. Datta,
Phys. Rev. Lett. \textbf{106}, 130503 (2011).





\bibitem{Schmidtrankofmixedstate} The Schmidt number of a mixed state $\rho$ on
$\mathbb{C}^{d}\otimes\mathbb{C}^{d'}$ is defined
by~[B. M. Terhal, P. Horodecki,
Phys. Rev. A \textbf{61}, 040301 (2000).]
\begin{eqnarray*}
S_r(\rho)=\inf\limits_{\mathcal{D}(\rho)}\max\limits_{\psi_i}S_r({\psi_i}),
\end{eqnarray*}
where $\mathcal{D}(\rho)=\{p_i,|\psi_i\rangle: \rho
=\sum_ip_i|\psi_i\rangle\langle\psi_i|\}$ is the set of all possible pure state ensembles
of the bipartite state $\rho$.

\bibitem{Schmidt} E. Schmidt,
Math. Ann. \textbf{63}, 433 (1907).

\end{thebibliography}

\end{document}